# Deep insights into the local structure of amorphous $Ta_2O_5$ thin films by X-ray pair distribution function analysis

Alberto Martinelli[1,*], Mauro Giovannini[2], Martina Neri [3] and Gianluca Gemme[3]

[1] *SPIN-CNR, Corso Perrone 24, 16152 Genova – Italy*

[2] *Department of Chemistry and Industrial Chemistry, University of Genova, Via Dodecaneso 31, 16146 Genova – Italy*

[3] *INFN, Sezione di Genova, Via Dodecaneso 33, 16146 Genova - Italy*

**Abstract**

Amorphous films of tantalum oxide ($Ta_2O_5$) are widely applied to build highly reflective mirrors used in interferometric gravitational wave detectors, such as the Laser Interferometer Gravitational Wave Observatory (LIGO). Despite a large number of studies, the structural properties of amorphous $Ta_2O_5$ at the local scale still deserve several further investigations. Such information is essential to any attempt to understand the properties of this important material and no property modelling efforts can be expected to yield reliable information until the local structure in the amorphous phase is better understood. In this paper we report the results obtained by analysing the synchrotron X-ray pair distribution functions of pure and Ti-doped amorphous $Ta_2O_5$ film, deposited by ion beam sputtering, as prepared and after a thermal treatment. As a result, it is found that 1) the arrangement of Ta atoms in amorphous $Ta_2O_5$ strongly resembles that characterizing the high-pressure $Z$-$Ta_2O_5$ polymorph, whereas the topological properties of O atoms resembles that observed in $\delta$-$Ta_2O_5$; 2) structural correlations in amorphous $Ta_2O_5$ start to vanish above ~ 5 Å and are completely suppressed for $r >$ 10 Å on account of disorder; 3) Ti substitution retains the short-range topological ordering characterizing the as prepared $Ta_2O_5$ amorphous film even after thermal treatment; conversely, pure $Ta_2O_5$ films undergoes a significant rearrangement of the local structure after thermal treatment.

---

[*] Corresponding author: alberto.martinelli@spin.cnr.it

# I. INTRODUCTION

Current interferometric gravitational wave detectors rely critically on the use of ultra-stable Fabry–Perot cavity arrangements [1,2,3,4]. In these systems, it is well known that a serious fundamental limit to the inherent performance is set by the Brownian motion associated with the ion-beam-sputtered cavity mirror coatings of amorphous oxide materials [5,6,7,8], with this noise source currently limiting, or imminently expected to limit, achievable experimental performance. The magnitude of this Brownian thermal noise is related to the mechanical dissipation factor of the coating materials. It is thus of major importance to determine the exact level of dissipation, and thus thermal noise, expected from specific coatings, understand the mechanism responsible for this dissipation and find methods of minimizing it. The mirrors in gravitational wave interferometers consist of multilayers of a low refractive index material alternating with a high refractive index material. Previous studies have shown that in the commonly-used coatings formed from alternating layers of silica ($SiO_2$) and amorphous-$Ta_2O_5$ ($a$-$Ta_2O_5$), the dissipation is dominated by the $Ta_2O_5$ component [9,10,11,12,13,14] and can be reduced by doping the $Ta_2O_5$ with Ti [15,16], although the mechanism responsible for the dissipation is still unclear. In this context, knowledge of the average local-scale structure of $a$-$Ta_2O_5$ is important to understand the properties of this material and to guide current property modelling efforts [17] and therefore many investigations were carried out.

Even though amorphous materials lack long-range symmetry, topological constraints are expected to hold to some extent, implying that $a$-$Ta_2O_5$ can share similar structural properties with one or more $Ta_2O_5$ crystalline polymorphs, at least on the local scale. Indeed, tens of polymorphic modifications are reported for crystalline $Ta_2O_5$ and, unfortunately, misnomers and improper structural descriptions are sometimes reported in literature giving rise to confusion in some cases.

EXAFS measurements ascertained that $a$-$Ta_2O_5$ exhibits a relatively well ordered local structure, resembling to some extent $\beta$-$Ta_2O_5$ [18], crystallizing in the orthorhombic *Pccm* space group [19]. A subsequent EXAFS investigation by Bassiri *et al.* [20] found that the distribution of the nearest O atoms around Ta appears in agreement with multiple structural models reported for $Ta_2O_5$ polymorphs, including the *Pccm* [19], the *Cmmm* [21] and the *Pbam* [22]. Nonetheless, these authors concluded that $a$-$Ta_2O_5$ is not a simple distortion of a crystalline structure, but, conversely, a significant rearrangement of the structural motif is involved. In particular, the coordination number of Ta in $a$-$Ta_2O_5$ is diminished down to 4 - 5, whereas it generally ranges around 6 – 7 [20] in the crystalline phases.

More reliable clues about O neighborhood were gained by $^{17}$O NMR analyses. Although no direct spatial information can be obtained by this technique, nuclei experiencing different fields give separate resonance lines in the NMR spectrum, thus providing an indication about the actual number of the different topological sites. Kim and Stebbins [23,24] found very similar $^{17}$O NMR spectra for both amorphous and crystalline (obtained by heating *a*-Ta$_2$O$_5$) samples. Moreover, they indexed the X-ray powder diffraction pattern of the crystalline phase according to the structural model of the *L*-Ta$_2$O$_5$ phase crystallizing in the *P2mm* space group [25]. In this context, it is worth to remind that also Oherlein *et al.* [26] indexed their diffraction patterns obtained by heating *a*-Ta$_2$O$_5$ at 900°C according to the same *L*-Ta$_2$O$_5$ structural model (these authors refer to this phase as the low-temperature *β*-Ta$_2$O$_5$ phase in their publication). Nonetheless, the structural model of the *L*-Ta$_2$O$_5$ phase is rather complex, constituted of 45 distinct atomic sites and a huge *b*-axis of ~ 40 Å [25]. All these features raise some doubts about the fact that the topological order of *L*-Ta$_2$O$_5$, although on the local scale, could resemble that of *a*-Ta$_2$O$_5$. In any case, Kim and Stebbins [23,24] ascribed the two peaks in their NMR spectra to a 2-fold and a 3-fold coordination sites of oxygen, respectively, by following diffraction outcomes and accordingly to the coordination environments found in the *L*-Ta$_2$O$_5$ crystal structure. Extremely similar diffraction patterns and $^{17}$O NMR spectra were obtained by Xu *et al.* [27] by analyzing Ta$_2$O$_5$ crystalline nanorods. Differently than Kim and Stebbins [24], they indexed the diffraction patterns according to the hexagonal *δ*-Ta$_2$O$_5$ polymorph (space group *P6/mmm*) isotypic with AlB$_2$ [28]. In this structural model O occupies one unique atomic position in a 6-fold trigonal prism coordination; hence this model does not comply with the bimodal distribution of topological oxygen sites provided by $^{17}$O NMR analyses. The diffraction and NMR results were finally reconciled by first-principles calculations of Fukumoto and Miwa [29] that redefined the O sub-structure in *δ*-Ta$_2$O$_5$ providing a *P6/mmm* structural model with both a 2-fold and a 3-fold coordination sites for oxygen, in agreement with NMR findings.

Some investigations based on the pair distribution function (PDF) analysis are also reported in literature. A first pioneering work based on a least-squares method analysis of laboratory X-ray diffraction data by Aleshina *et al.* [30] concluded that the short-range order of *a*-Ta$_2$O$_5$ films is similar to that characterizing the *β*-Ta$_2$O$_5$ polymorph reported by Holser [31], crystallizing in the *Pmmm* space group [32]. A PDF from *a*-Ta$_2$O$_5$ films was also obtained by radially averaged electron diffraction data and fitted by using a Reverse Monte Carlo refinement [33,34,35]. As a result, the nearest Ta-O pairs were found in good agreement with several polymorphs and from the refined model the coordination numbers for Ta and O were estimated to be ~ 6.5 and ~ 2, respectively. Moreover, no structural order was found beyond ~ 4 Å. A PDF study based on grazing incident synchrotron X-ray radiation by Shyam *et al.* [36] revealed that actually some degree of ordering propagates up to ~ 10 Å in $a$-$Ta_2O_5$ thin films.

Hereinafter we present high resolution synchrotron PDF data of pure and Ti-substituted $a$-$Ta_2O_5$ thin films; in particular as-deposited and air-annealed samples were analyzed. Data were collected by means of a normal-incidence synchrotron X-ray beam experimental set-up. This experimental configuration enables to acquire data at higher scattering vector Q values, thus obtaining a better resolution of the experimental PDF.

In this work we argue that the Ta-substructure displays a short-range atomic arrangement in $a$-$Ta_2O_5$ films, resembling to some extent that observed in the crystalline Z-$Ta_2O_5$ phase [37], whereas the O atoms are topologically distributed between 3-fold coordinated (intra-layer) and 2-fold coordinated (inter-layer) sites. Moreover, Ti substitution quenches the local structure of $a$-$Ta_2O_5$, that is otherwise significantly rearranged by thermal treatment.

## II. EXPERIMENTAL METHODS

$a$-$Ta_2O_5$ samples are 2 μm-thick amorphous films deposited by ion beam sputtering (IBS) on 200 μm-thick amorphous $SiO_2$ substrates (1 inch diameter). The films were produced by Laboratoire des Matériaux Avancés (LMA) in Lyon (France) through double ion beam sputtering technique; some of them underwent a subsequent annealing treatment. Two kinds of samples were prepared, a pure $Ta_2O_5$ and a Ti-substituted films. Some of these samples then underwent a thermal treatment, by heating at 500 °C for 10 hours in air. Hereinafter, the samples are labelled as Ta_NA ($a$-$Ta_2O_5$ not annealed), Ta_an ($a$-$Ta_2O_5$ annealed), Ti_NA (Ti-substituted $a$-$Ta_2O_5$ not annealed) and Ti_an (Ti-substituted $a$-$Ta_2O_5$ annealed).

In order to evaluate the Ti-content, the Ti-substituted sample was first embedded in cold-setting epoxy, then underwent metallographic polishing and finally the [Ta]/[Ti] ratio was determined by SEM-EDS analyses. As a result the [Ta]/[Ti] ~ [78]/[22] ratio remains roughly constant in the whole film thickness; remarkably, a few amounts of Ar (~ 3% at.) is also detected, implanted during film deposition.

The pair distribution function (PDF) analysis is a well-established technique for studying the local structure of amorphous and crystalline materials as well. This function is defined in the direct space and represents the probability to find a given atomic pair separated by a distance *r* in a compound, regardless of its crystalline or amorphous state. Moreover, a structural model can be developed and fitted within different distance ranges of the experimental PDF data, thus providing a direct method for both testing the reliability and goodness of the hypothetical structure and the deviation from the average structure at different length scales. At this scope, diffraction data were collected at room temperature at the old ID31 beamline of the European Synchrotron Radiation Facility in Grenoble (F), using a X-ray beam with a wavelength $\lambda = 0.15895$ Å. Data from an amorphous $SiO_2$ substrate were collected to subtract the substrate scattering; the calibration of the detector and beam centering were carried out using a $CeO_2$ sample. The experimental set-up used for data collection is sketched in Figure 1, with the film surface perpendicular to the scattering beam, and is similar to that described by Jensen *et al.* [38], where is named tfPDF (thin film PDF) method. In our experiment the incident X-ray beam was first scattered by the thin film and subsequently by the substrate. This experimental set-up is advantageous in respect of the grazing incidence technique adopted in previous works [36,39], since that the set of problems related to angular-dependent corrections for the penetration depth and the substrate scattering are suppressed. Moreover, it allows to collect data at larger scattering vector **Q** values, thus improving the quality of the final PDF function.

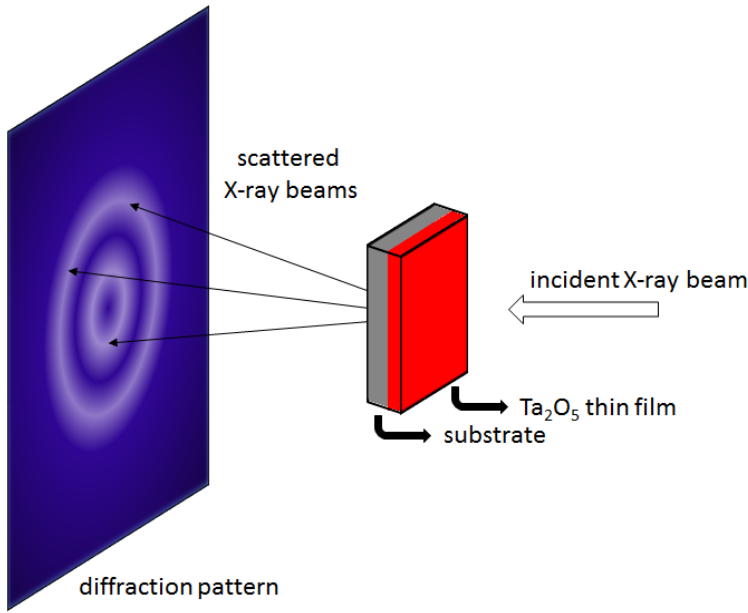

Figure 1: Sketch of the experimental set-up used for data collection at the old ID31 beamline of ESRF; the X-ray beam first interacts with the amorphous $Ta_2O_5$ film and afterwards with the silica substrate.

The $G(r)$ function (the reduced atomic PDF) is calculated by integrating the total-scattering structure function $S(\mathbf{Q})$ obtained from the scattered intensity $I(\mathbf{Q})$:

$$G(r) = \frac{2}{\pi} \int_0^\infty Q[S(\mathbf{Q}) - 1] \sin(Qr) \mathrm{d}Q$$

Inasmuch the integration for $G(r)$ calculation is ideally extended up to an infinite value of the scattering vector **Q**; as a consequence, more detailed structural information and minimization of spurious effects are gained with the increase in the available maximum value of **Q** ($Q_{max}$) in the experimental data $I(\mathbf{Q})$. In particular, the PDF resolution $dr$ is dominated by the $Q_{max}$ used in the Fourier transform as [40]:

$$dr = \frac{\pi}{Q_{max}}$$

Reduction of the total scattering data to obtain $G(r)$ was done by the PDFgetX3 software [41] using $Q_{max} = 26.0$ Å$^{-1}$ (by comparison, in previous grazing incidence X-ray PDF studies $Q_{max}$ is 20 Å$^{-1}$ [36,39]) and $r_{poly} = 0.9$ Å (default value). The $G(r)$ functions of the inspected thin films were obtained by using the data from the pure amorphous SiO$_2$ substrate as backgroundfile in PDFgetX3.

Full-profile fitting of the $G(r)$ function was carried out by refining a periodic structural model, a method suited for compounds which are close to a crystalline state. At this scope the PDFgui software was used [42]. Instrumental parameters ($Q_{damp}$ and $Q_{broad}$) were refined using a standard CeO$_2$ sample and fixed during the subsequent fittings of the $a$-Ta$_2$O$_5$ data; moreover, the structural coherence length was refined using the *spdiameter* parameter in order to account for the amorphous nature of the investigated samples. In the end, the following parameters were refined for profile fitting: the scale factor; the unit cell parameters; the atomic positions not constrained by symmetry; the isotropic atomic displacement parameters; the *spdiameter* parameter.

## III. RESULTS

Crystalline solids exhibit long-range periodicity, which is a symmetric arrangement of atoms in space, producing a regular repetition of the number of bonds, bond lengths and angles at equivalent crystallographic positions (structural constraints). Despite the lack of long-range symmetry, there are topological constraints at the local scale in amorphous materials as well. Then, structural constraints can display similar values in compounds found as both amorphous and crystalline phases. Indeed, in an amorphous phase the structural constraints can be regularly repeated only within a narrow distance range and topological order can develop only at a short-range scale. The crystal chemical properties of Ta (connected to its coordination number and oxidation state) are then expected to possibly produce similar atomic arrangements at the short-range scale in $a$-Ta$_2$O$_5$ as those observed in the homologous crystalline materials. In this context, the local structure of a material can be described by the distribution of the interatomic distances, whatever its degree of crystallinity. From this point of view, the $G(r)$ function is effective: being a histogram of interatomic distances, this function describes the average structure and provides an analytical tool for a reliable and efficient comparison among the structures of crystalline and amorphous materials at the short- to medium-range scale.

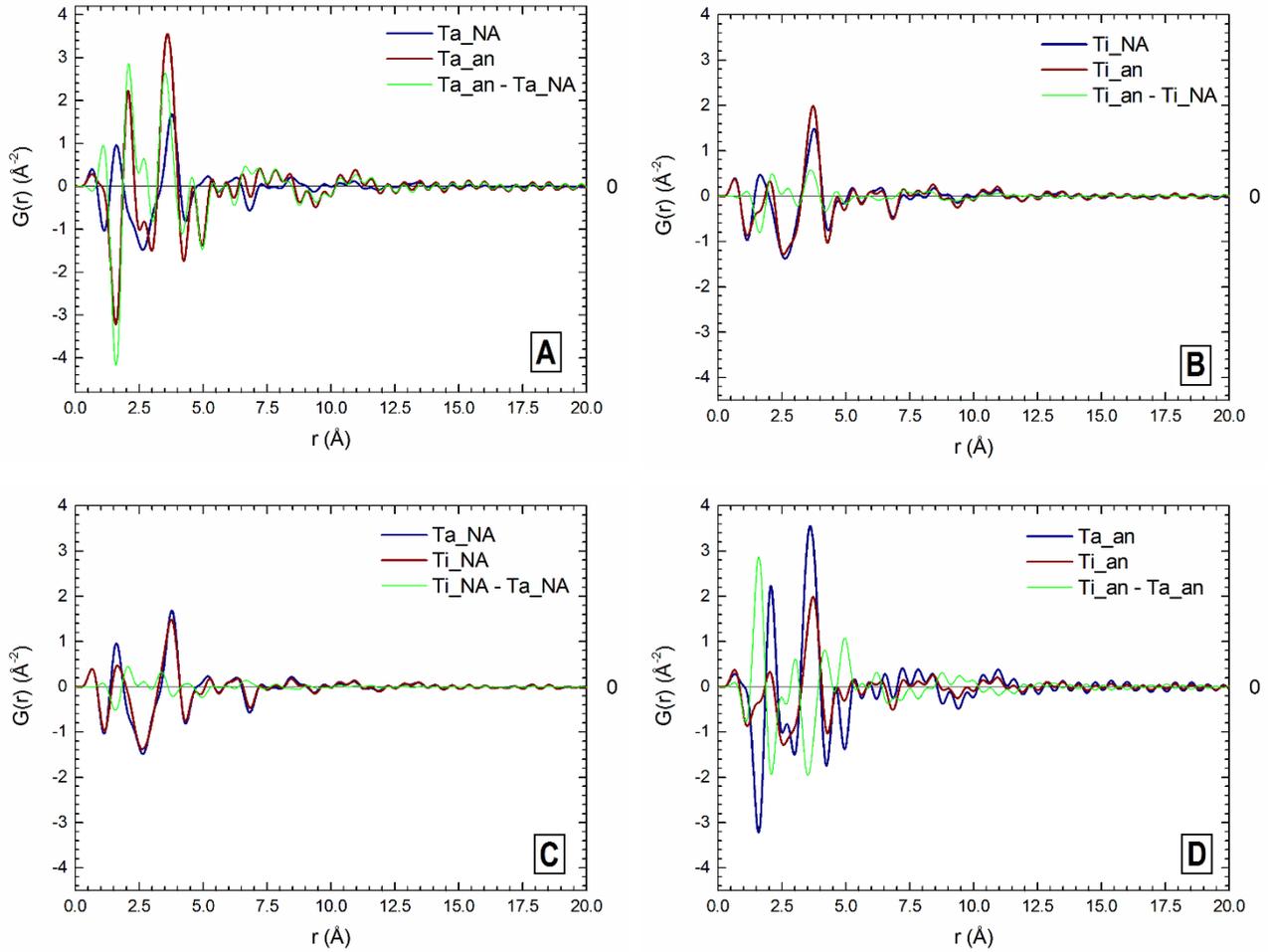

Figure 2: Superposition of the $G(r)$ functions obtained for the different $Ta_2O_5$ thin films and corresponding difference curves.

Figure 2 shows the comparison between pairs of experimental $G(r)$ functions obtained for the inspected $a$-$Ta_2O_5$ thin films after subtraction of the $SiO_2$ substrate contribution. At a first glance, these functions are consistent with those reported in literature [33,34,35,36,39], although more subtle details in terms of peak shape can be detected. The Ta ($Z = 73$) contribution to the $G(r)$ function dominates over that of O ($Z = 8$) as the experimental data were collected by using X-ray radiation. Hence, the strongest peak is centered between 3 and 4 Å and corresponds to Ta – Ta distances, whereas the first peak at ~ 1.5 – 2.5 Å corresponds to the first-neighbor Ta – O distance. A very weak contribution of the O – O pair is noticed at ~ 2.7 Å only in the Ta_an sample, suggesting an increasing of structural order induced by the thermal treatment. Moreover, the contribution of both Ta-Ta and Ta-O pairs in this sample is still effective for the peak at ~ 4.5 Å in this sample, indicating a persistence of a notable degree of structural ordering up to this range scale. This is in agreement with Hart *et al.* [43], that observed changes in the medium range ordering of the atomic structure of amorphous tantala after thermal annealing.

Generally speaking, the amplitude of the oscillations observed in the $G(r)$ function gives a direct measure of the structural coherence, whereas no peak is observed at a distance exceeding structural correlations. In our samples the correlations between the atomic positions decrease with the increase of the inter-atomic separation. Roughly three regions can be identified in the distance scale as the structural coherence in *a*-$Ta_2O_5$ is progressively lost on account of disorder: 1) a low-range region ($r < 4.5$ Å), where topological order is quite well developed; 2) an intermediate-range region (4.5 Å $< r < 10$ Å), where the amplitude of the oscillations decreases, thus indicating that structural order is noticeably and progressively frustrated; 3) a long-range region ($r > 10$ Å), where the $G(r)$ function vanishes (ripples above ~ 10 Å are data noise) pointing to a definitive suppression of the structural correlations.

Thermal and/or static disorder gives rise to a distribution of the interatomic distances, inducing a broadening of the PDF peak. As a result, the width of the peaks of the $G(r)$ function is determined by the width of the distribution of the interatomic distances between a given pair of atoms. The annealing treatment increases the symmetry of the distribution of both Ta-O (peak at ~ 1.5 - 2.5 Å) and Ta-Ta pairs (peak at ~ 3.0 - 4.0 Å), as evidenced by comparing the Ta_NA and Ta_an samples (constituted of pure *a*-$Ta_2O_5$; Figure 2, panel A). Ti-substitution induces a slight disordering in (Ta,Ti)-O bond distance distribution, whereas the arrangement of metal atoms results almost unaffected by chemical substitution (Figure 2, panel C). Indeed thermal annealing rearranges the distribution of the (Ta,Ti)-O bonds, but to a much less extent in respect to the pure *a*-$Ta_2O_5$ samples (Figure 2, panel B and D). These observations differ with what observed in annealed multilayer coatings, where the radial distribution functions obtained by electron diffraction evidences that homogeneity grows with the increase of Ti content (increase in height and decrease in width of the peaks) [33]. Conversely our data indicate that Ti substitution mainly produces a slight displacement of O atoms, without relevant changes in the metal sub-structure. This is the reason why the difference curves of panels A and D in Figure 2 are almost specular. It can be thus concluded that Ti-substitution freezes to a large extent the atomic distribution of the as-prepared thin *a*-$Ta_2O_5$ films, even after thermal treatment. Conversely, the pure *a*-$Ta_2O_5$ film undergoes a significant rearrangement of its local structure after annealing.

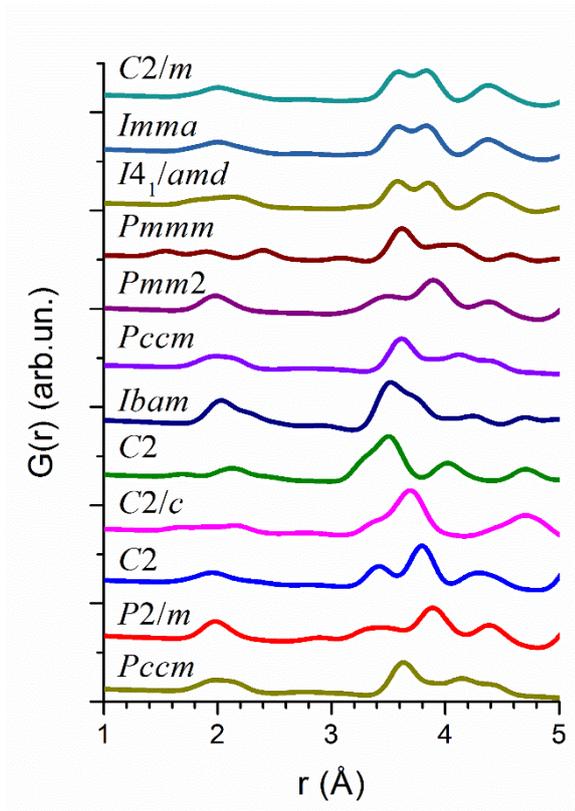

Figure 3: Comparison of the normalized $G(r)$ functions calculated for selected polymorphic structures of $Ta_2O_5$ (calculated from available structural data); the space group symbol for each structure is indicated (note that 2 different $C2$-type structures are reported in literature).

Crystalline $Ta_2O_5$ displays several polymorphic modifications. Figure 3 shows a comparison of the $G(r)$ functions obtained by using the available crystallographic data for selected polymorphs (calculations were carried out by applying the parameters describing the experimental instrumental resolution, as obtained by the standard $CeO_2$ sample). Only short range-data are reported (up to 5 Å) for comparing the local structure of the inspected amorphous samples with those of the crystalline phases.

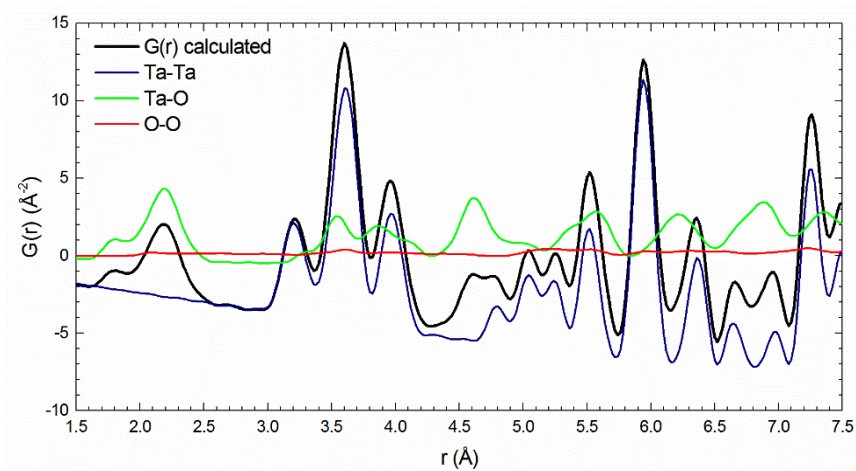

Figure 4: Contribution of the different atomic pairs to the total $G(r)$ function; calculations were carried out using the structural data of the $C2$ polymorph of $Z$-$Ta_2O_5$.

Selected structural data of these polymorphs were used as starting structural models for fitting the PDF data collected for $a$-$Ta_2O_5$ thin films. The monoclinic $C2$ structural model of the $Z$-$Ta_2O_5$ phase [37] provides the best result (See Supplemental Material at [44] for PDF data fitting), taking into account the goodness of fitting and the reliability of the refined structure (on the basis of crystal chemical parameters, such as next nearest Ta-O bond lengths). In this context it is worth to note that this *C*2 polymorphic modification is characterized by the lowest unit cell volume on account of its rather simple atomic arrangement. Nonetheless, this result should not be overrated, since that the determination of the exact location of O atoms is extremely challenging, because of the relatively low value for the X-ray scattering length characterizing this atomic species. Figure 4 shows an enlarged view of the $G(r)$ function for Z-Ta$_2$O$_5$, calculated for a $10a \times 10b \times 10c$ structure, using the structural parameters of the Ta_NA sample obtained by applying the *C*2 structural model for data fitting and assuming $Q_{max}$ = 30 Å. The calculated partial $G(r)$ functions containing only contributions of the Ta-Ta, Ta-O and O-O pairs are superposed. The weights for PDF calculation are 7.545, 0.827 and 0.091 for the Ta-Ta, Ta-O and O-O pairs, respectively. It is evident that only Ta-O pairs contribute to the peak at ~ 2.0 Å, whereas the broad peak at ~ 3.5 Å is dominated by Ta-Ta pairs. Conversely, the contribution of the O-O pairs is almost negligible along the whole r distance, clearly indicating that the oxygen positions refined by using the Z-Ta2O5 structural model must be cautiously considered. On the other hand, the distribution of Ta atoms obtained after the fitting of the PDF can be reliably evaluated.

It is thus instructive to compare the Ta-substructure of the Z-Ta$_2$O$_5$ polymorph with those characterizing other structural models that were proposed in previous works (hereinafter the reader should be aware of the fact that the notation for the different crystalline Ta$_2$O$_5$ phases is quite confusing and sometimes the polymorphic modifications cited in the original works were later redefined).

In their pioneering investigation, Aleshina *et al.* [30] report that the short-range order in *a*-Ta$_2$O$_5$ shares similar features with that characterizing the crystalline *β*-Ta$_2$O$_5$ polymorph, but refer to two different structural models for this phase. In the first case the reference structural model is orthorhombic (space group *Pmm*2 – n° 25) with lattice parameters (Å) $a$ = 6.198, $b$ = 40.290, $c$ = 3.888 [25]. The huge dimension of the *b* parameter brings into question that *a*-Ta$_2$O$_5$ could share a widespread similarities with this structure. This conclusion is corroborated by the fact that the metric of the unit cell reported by Aleshina *et al.* [30] for their sample is remarkably different ($a$ = 6.20 Å, $b$ = 3.66 Å, $c$ = 3.89 Å). In the second case, the considered structural model is again orthorhombic, but pseudo-hexagonal. In their work Aleshina *et al.* [30] refer to the attempts of Holser [45], reporting that *β*-Ta$_2$O$_5$ has the *α*-U$_3$O$_8$ structure type with lattice parameters (Å) $a$ = 6.20, $b$ = 3.67, $c$ = 3.90. Indeed, the *α*-U$_3$O$_8$ structure has a very large *b* parameter (~20 Å) and actually no Ta$_2$O$_5$ polymorph was subsequently confirmed to crystallize with the *α*-U$_3$O$_8$ structure. Nonetheless, the cell parameters listed by Holser [45] are almost exactly the same measured by Lehovec [32] in the *Pmmm* (n° 47) *β*-Ta$_2$O$_5$ polymorph and are consistent with those reported by Aleshina *et al.* [30] for *a*-Ta$_2$O$_5$. Attempts to fit our data with the *Pmmm* structural model were carried out, but the goodness of fitting worsens in comparison with the *C*2 one (See Supplemental Material [44] for PDF data fitting). Nonetheless it is worth to remark that Aleshina *et al.* [30] found a similar short range order for both amorphous Ta$_2$O$_5$ and Nb$_2$O$_5$ and that the Z-Ta$_2$O$_5$ phase is exactly isotypic with Nb$_2$O$_5$.

The $^{17}$O NMR analyses coupled with X-ray diffraction [23,24,27] give important clues in outlining the topological properties of O atoms, indicating that O atoms are distributed in 2- and 3-fold coordination sites. It is particularly tempting to assume that these properties are very similar in the *δ*-Ta$_2$O$_5$ polymorph [29] (growth by annealing amorphous Ta$_2$O$_5$ [24]) and the *a*-Ta$_2$O$_5$ phase itself, on account of $^{17}$O NMR results.

Remarkably, the Ta atoms are distributed according to a hexagonal lattice in both *β*-Ta$_2$O$_5$ (*Pmmm* structure) and *δ*-Ta$_2$O$_5$ (*P*6/*mmm*), whereas the Ta substructure can be viewed as a distorsive derivative of a primitive hexagonal structure in the Z-Ta$_2$O$_5$ polymorph (*C*2), as sketched out in Figure 5. Hence a novel structural model can be hypothesized, characterized by a Ta substructure similar to that of the Z-Ta$_2$O$_5$ polymorph, but with O atoms distributed in 2-fold (in inter-layer sites) and 3-fold coordination (in intra-layer sites) as in *δ*-Ta$_2$O$_5$. On the other hand, the amount of structural information contained in the PDF data is intrinsically limited by the fact that atomic correlations start to vanish for r > 4.5 Å. This implies that the number of free parameters (and hence the number of independent atoms) introduced in the structural model must be consequently limited. Given this, we developed a *P*1 model where in the first step Ta atoms are arranged in hexagonal layers, but without being stacked along a main axis as in the *β*-Ta$_2$O$_5$ and *δ*-Ta$_2$O$_5$ phases (Figure 5).

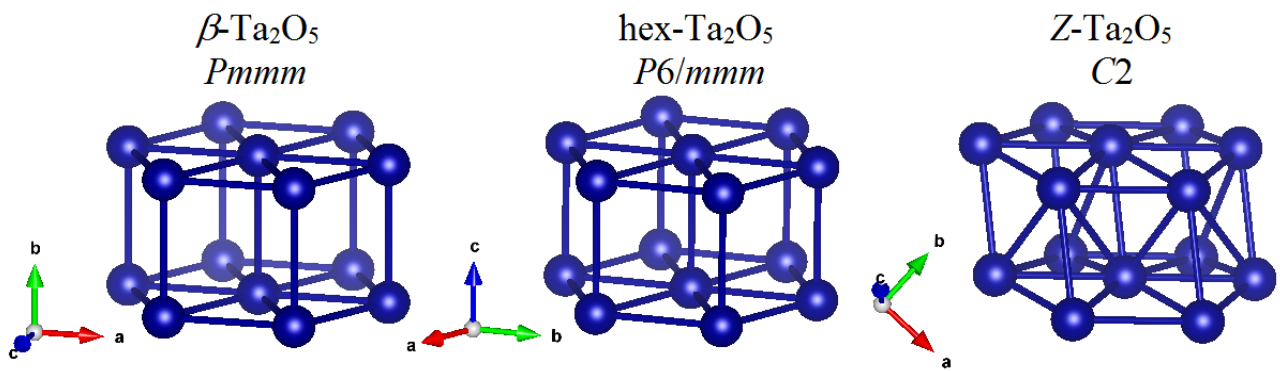

Figure 5: Comparison of the Ta substructure in *β*-Ta$_2$O$_5$ (*Amm*2), hexagonal Ta$_2$O$_5$ (*P*6/*mmm*) and Z-Ta$_2$O$_5$ (*C*2) polymorphs.

In this way it is possible to reproduce the shifted stacking of the hexagonal layers charactering the $Z$-$Ta_2O_5$ phase that comes out as a characteristic feature by the PDF analysis of the inspected thin films. Unfortunately, this model can not reproduce also the slight displacements of Ta atoms above and below the ideal plane that are actually found in the $Z$-$Ta_2O_5$ phase (Figure 5). At this scope a more complex structural model would be required, containing a larger number of both independent Ta and O atoms (i.e. the number of independent free parameters would notably increase). In the second step O atoms are distributed in 2-fold coordination (inter-layer sites), directly intervening between two neighboring inter-layer Ta atoms, and in 3-fold coordination (intra-layer sites), similarly to what observed in the $\delta$-$Ta_2O_5$ phase (Figure 6).

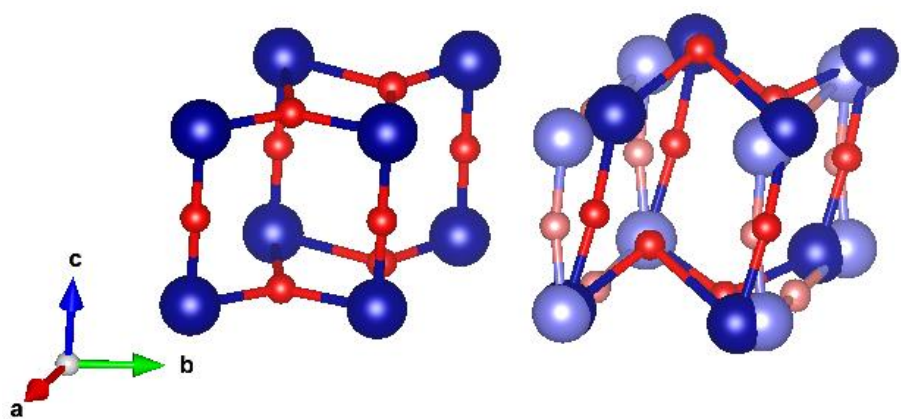

Figure 6: Panel on the left: Distribution of the 2- and 3-fold coordination sites of O atoms in the $\delta$-$Ta_2O_5$ polymorph. Panel on the right: Superposition of the local structure of $a$-$Ta_2O_5$ in the Ta_an and Ta_NA samples (light and dark coloured atoms, respectively).

This triclinic $P1$ structural model provides a reasonably good fitting of the $G(r)$ functions (Figure 7; Table 1 lists the resulting unit cell parameters; (See Supplemental Material [44] where all the refined structural data are listed) and it is able to reproduce the main features characterising the experimental data. The observed differences can be ascribed to the intrinsic disorder of the amorphous phase that cannot be accounted in a crystalline structural model.

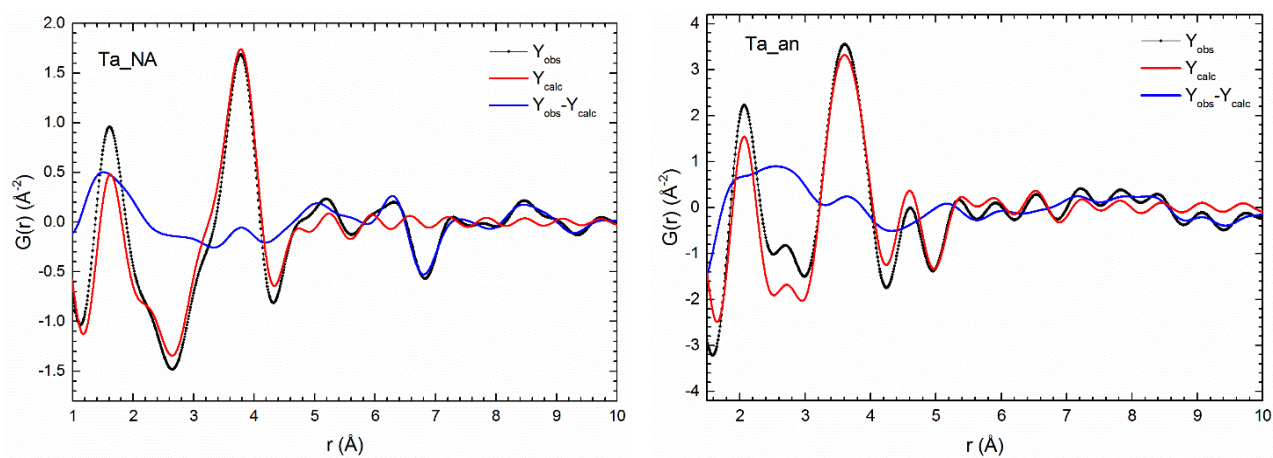

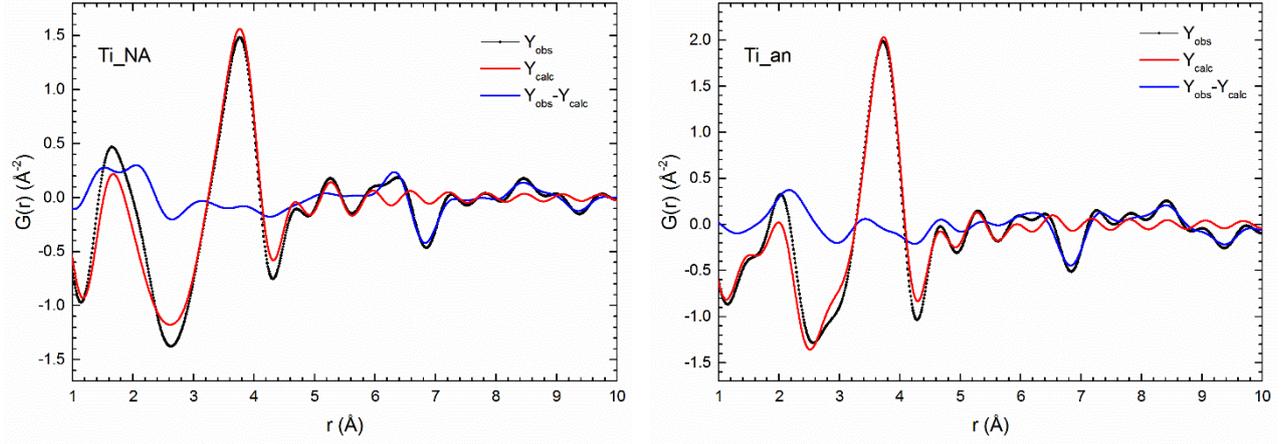

Figure 7: The $G(r)$ functions of the inspected samples fitted with the $P1$ structural model.

Data fitting confirms the distinctive properties of the Ta_an samples from those of the other samples, inasmuch it is characterized by notably different triclinic angles as well as a structural coherence length that develops significantly further. Indeed, the extent of the structural coherence in the real space can be estimated by assuming an isotropic distribution of pair correlations from the spherical shape factor (*spdiameter* parameter; Table 1), since that the $G(r)$ function gets damped at higher distances as the structural coherence progressively vanishes. As a result the Ta_an sample is characterized by a significant larger structural coherence, whereas Ti-substitution appears to quench the crystallization even after thermal annealing.

Table 1: Unit cell parameters obtained by fitting the $G(r)$ functions for amorphous thin films with a periodic $P1$ structural model (calculation up to 10 Å); the goodness of fitting is evaluated by $R_w$.

|  | **Ta_NA** | **Ta_an** | **Ti_NA** | **Ti_an** |
|---|---|---|---|---|
| $a$ (Å) | 3.63(6) | 3.49(2) | 3.59(6) | 3.57(6) |
| $b$ (Å) | 3.9(2) | 3.31(2) | 3.84(2) | 3.50(7) |
| $c$ (Å) | 3.41(6) | 3.89(4) | 3.48(6) | 3.81(9) |
| $\alpha$ (deg) | 93(4) | 80.9(7) | 90(4) | 93(3) |
| $\beta$ (deg) | 116(2) | 121.1(6) | 115(4) | 118(3) |
| $\gamma$ (deg) | 116(3) | 112.1(6) | 118(8) | 113(2) |
| *volume* (Å$^3$) | 36.8±2.3 | 35.6±0.6 | 37.3±4.0 | 36.6±2.2 |
| *spdiameter* | 6.4(5) | 8.8(6) | 6.5(6) | 6.9(6) |
| $R_w$ (%) | 33.18 | 33.75 | 26.88 | 25.12 |

The Figure 6, panel on the right, shows the atomic arrangements in the Ta_an sample (atoms drawn with darker colors) and how atoms are displaced in the Ta_NA sample (atoms drawn with lighter colors). In the Ta_an sample, O sites with 2-fold (inter-layer sites) and 3-fold coordination (intra-layer sites) are observed similarly as in the $\delta$-Ta$_2$O$_5$ phase. In particular, the diffuse presence of 3-fold coordinated intra-layer sites determines a 3-dimensional character of the local structure. This 3-dimensional character is also corroborated by the peak at 4.2-4.9 Å, just originated by Ta-O intra- and inter-layer bond distances, thus confirming the development of a substantial structural coherence involving both intra- and inter-layer O sites.

Although the atomic arrangements in the Ta_an sample can be still considered a distorted derivative of that observed in the $\delta$-$Ta_2O_5$ phase, more striking differences characterize the Ta_NA sample (and the Ti-substituted ones as well). In particular, the intra-layer site tends to reduce (on average) their coordination number in the Ta_NA sample on account of the displacement of the O atom in an intervening position between two neighboring intra-layer Ta atoms. Such a transformation reflects an increasing of 2-dimensional character of the local structure of the Ta_NA sample (and Ti-substituted samples as well) in comparison with the Ta_an sample. This result is consistent with both $^{17}$O NMR results [24], where a notable reduction of the 3-fold coordination sites where observed by comparing crystalline and amorphous $Ta_2O_5$, as well as previous PDF analyses [36], where a lack of 3-dimensional order was argued in both pure and Zr-substituted $a$-$Ta_2O_5$.

Indeed, a systematic comparison of the PDF functions reported in literature [33,34,3536,39] reveals a fair agreement with our data (all the investigated samples were synthesized by ion beam sputtering, as our specimens). All these data are characterized by two main strong peaks, in between 1.5-2.0 Å (first-neighbor Ta – O pair) and 3.0-4.0 Å (first-neighbor Ta – Ta pair), respectively, as in our PDF functions.

Consistently with our data, the $G(r)$ function obtained by radially averaged electron diffraction data for amorphous $Ta_2O_5$ coatings reveals that order starts to drop for $r > 4$ Å [33,34,35]. Also the $G(r)$ function obtained by synchrotron X-ray data in a grazing incidence geometry by Shyam *et al.* [36] is consistent with our data. The peaks observed by these authors are well reproduced in our data (taking into account the intrinsic variations characterizing the different samples as well as the different experimental conditions) and, as in our data, structural correlations are completely suppressed for $r > 10$ Å in the amorphous sample. Moreover these authors underline a striking similarity in the $G(r)$ functions of the amorphous and crystalline samples below ~ 4 Å, confirming that the topological order characterizing the crystalline phase is maintained at the short local scale even in the amorphous phase. It is interesting to observe that the $G(r)$ function of the crystalline sample is characterized by a weak, but well defined peak at 2.8 Å produced by O-O pairs and its suppression in the amorphous sample is ascribed to the disordering of the coordination polyhedral [36]. Actually, in our data this peak is clearly observed in the Ta_an data and is reproduced in our $P$1 structural model (see Figure 2 and Figure 7), confirming that the annealing treatment in the pure $a$-$Ta_2O_5$ coating determines an increase of the local structural order. This conclusion is in fair agreement with previous electron diffraction investigations [43], where an increasing of order was detected in $a$-Ta$_2$O$_5$ after annealing at 600 °C (the O-O pair contribution can be also detected by a closer inspection of Fig 3 in ref. [43] as a distinct shoulder growing around ~0.38 Å$^{-1}$ (~2.62 Å) after annealing at 600°C).

Conversely, the weak peak located at ~ 4 Å in the amorphous sample and ascribed in the crystalline sample to the regular stacking along the $c$-axis [36] is suppressed. Again, this occurrence is in perfect agreement with our model for the Ta-substructure, where Ta atoms are arranged according to a distorted hexagonal arrangement with displaced layers stacked along the c-axis (Figure 5 and Figure 6).

The rough retention of the as-deposited structure of $a$-Ta$_2$O$_5$ by Ti-substitution, even after thermal treatments, is consistent with previous results obtained on Zr-substituted $a$-Ta$_2$O$_5$, where crystallization hindering were ascribed to slight differences in the Ta-O and Zr-O bond distances [39, 46].

As a conclusion, all these data corroborates our findings, that thus constitute a relevant step forward in the comprehension of the local structure of amorphous Ta$_2$O$_5$.

## V. CONCLUSIONS

The local structure of pure and Ti-doped amorphous Ta$_2$O$_5$ films (both as deposited and after annealing) was investigated by means of synchrotron X-ray pair distribution analysis. The observed short-range arrangement of the Ta substructure strongly resembles that observed in the high-pressure $Z$-Ta$_2$O$_5$ polymorph, whereas O atoms are distributed in 2- and 3-fold coordination sites similarly as in the hexagonal $\delta$-Ta$_2$O$_5$ polymorph. The local structure of $a$-Ta$_2$O$_5$ can be thus represented as a distorted derivative of that characterizing the $\delta$-Ta$_2$O$_5$ polymorph, where atomic pairs progressively loose structural coherence above ~ 4.5 Å and atomic correlations are completely suppressed above ~ 10 Å. Remarkably, Ti-substitution slightly affects the local atomic arrangement, but rather quenches the amorphous structure of the as-prepared pure amorphous Ta$_2$O$_5$ even after thermal treatment, likely on account of the (Ta,Ti)-O bond distance fluctuations. Conversely, thermal annealing significantly affects the local structure of the pure amorphous Ta$_2$O$_5$ films, also increasing the structural coherence to a larger distance. As a consequence, the prevailing 2-dimensional character of the as-deposited $a$-Ta$_2$O$_5$ gain a 3-dimensional nature after annealing, even though the role played by the structural strain induced by the substrate in this process merits further investigation. These results could represent a fundamental step towards a better understanding of the dissipation mechanism affecting interferometric gravitational wave detectors and improving their performance.

**ACKNOWLEDGMENTS**

Authors acknowledge dr. A. Poulain for her kind support and experimental assistance during the data collection at the ID31 beam line of ESRF (proposal HC-2315). The authors gratefully acknowledge C. Michel (Laboratoire des Matériaux Avancés, Institut de Physique des 2 Infinis de Lyon) for helpful discussion on sample preparation.